\documentclass[10pt,journal,compsoc]{IEEEtran}
\usepackage{cite}
\usepackage{amsmath,amssymb,amsfonts}
\usepackage{algorithmic}
\usepackage{graphicx}
\usepackage{textcomp}
\usepackage{xcolor}
\usepackage{booktabs} 
\usepackage{multirow}
\long\def\comment#1{}
\begin{document}
	
	\title{AppQ: Warm-starting App Recommendation\\ Based on View Graphs}
	
	\author{Dan Su, Jiqiang Liu, Sencun Zhu, Xiaoyang Wang, Wei Wang, Xiangliang Zhang
		
		\IEEEcompsocitemizethanks{
			\IEEEcompsocthanksitem D. Su, J. Liu and W. Wang are with Beijing Key Laboratory of Security and Privacy in Intelligent Transportation, Beijing Jiaotong University, Beijing, China.\protect\\
			E-mail: \{sudan1, jqliu, wangwei1\}@bjtu.edu.cn
			
			\IEEEcompsocthanksitem S. Zhu is with Department of Computer Science and Engineering, The Pennsylvania State University, PA, USA\protect\\
			E-mail: sxz16@psu.edu
			
			\IEEEcompsocthanksitem X. Wang is with Beijing Key Laboratory of Traffic Data Analysis and Mining, Beijing Jiaotong University, Beijing, China\protect\\
			E-mail: shawnwang@bjtu.edu.cn

			\IEEEcompsocthanksitem X. Zhang is with CEMSE Division, King Abdullah University of Science and Technology (KAUST), Thuwal, Saudi Arabia\protect\\
			E-mail: xiangliang.zhang@kaust.edu.sa}
		
		%\thanks{Manuscript received in 2020}
		
	}

	% The paper headers
	\markboth{Journal of \LaTeX\ Class Files,~Vol.~14, No.~8, August~2015}%
	{Shell \MakeLowercase{\textit{et al.}}: Bare Demo of IEEEtran.cls for Computer Society Journals}
	
	\IEEEtitleabstractindextext{%
		\begin{abstract}
			Current app ranking and recommendation systems are mainly based on user-generated information, e.g., number of downloads and ratings. However, new apps often have few (or even no) user feedback, suffering from the classic cold-start problem. How to quickly identify and then recommend new apps of high quality is a challenging issue. 
			Here, a fundamental requirement is the capability to accurately measure an app's quality based on its \textit{inborn} features, 
			rather than user-generated features. Since users obtain first-hand experience of an app by interacting with its views, we speculate that the inborn features are largely related to the visual quality of individual views in an app and the ways the views switch to one another. 
			In this work, we propose AppQ, a novel app quality grading and recommendation system that extracts \textit{inborn} features of apps based on app source code. 
			In particular, AppQ works in parallel to perform code analysis to extract app-level features as well as dynamic analysis to capture view-level layout hierarchy and the switching among views.  Each app is then expressed as an attributed view graph, which is converted into a vector and fed to classifiers for recognizing its quality classes.
			Our evaluation with an app dataset from Google Play reports that AppQ achieves the best performance with accuracy of 85.0\%. This shows a lot of promise to warm-start app grading and recommendation systems with AppQ.
		\end{abstract}

		\begin{IEEEkeywords}
			Android, Warm-start, App Recommendation, App View Graph
	\end{IEEEkeywords}}

	\maketitle

	\IEEEdisplaynontitleabstractindextext
	\IEEEpeerreviewmaketitle

	\IEEEraisesectionheading{\section{Introduction}\label{sec:introduction}}
	
	\IEEEPARstart{A}{ndroid} is currently the most popular mobile operating system with a global market share of 75.39\% \cite{market-share}. The open nature of Android has led to a nearly exponential growth in mobile apps in recent years. The number of available apps in the Google Play Store reached 2.9 millions in December 2019, after surpassing 1 million apps in 2013 \cite{num-of-app-googleplay}. On the one hand, the dazzling number of apps provide convenience and enrich users' life; on the other hand, the great quantity of apps under each category makes it difficult for users to make choices. It is simple for users to download apps by several clicks; however, it is not so simple to find high-quality apps. The remarkable growth of Android app markets increasingly requires recommendation methods that can rank and recommend apps \textit{automatically} and \textit{accurately}.  
	
	To make a quick and effective decision on selecting apps for download, users commonly pay attention to the ratings and number of downloads, which are shown on the introduction page of an app. Few would further refer to reviews written by previous users. If the reviews are positive, new users would have a higher confidence to download the app. Although it is not completely known how apps are ranked, there are strong evidences that the number of downloads, ratings and reviews play the major roles. While such ranking strategies work to some extent, they bear an inherent drawback. That is, newly published apps suffer from the cold-start problem because of the lack of user generated data on them. Although most app markets run a vetting process, currently the process mainly concerns with security violation, rather than the quality of an app. App markets have little to leverage when ranking or recommending new apps. Given the giant size of app markets, some high-quality apps may not get a good opportunity to be recognized, which can easily discourage the developers from making more excellent apps. Therefore, purely based on user generated data to grade apps is one-sided.
	
	Previous work on grading and recommendation systems mainly focused on user generated information. For instance, DroidVisor \cite{RustgiFRM17} clustered apps based on ratings, downloads and reviews. They are more concerned about user' comments that may be misleading and indiscriminate rather than the real quality of apps. A few more app ranking related systems, such as RankMyApp \cite{Rankmyapp} and Appfigures  \cite{Appfigures}  offer app store ranking rather than optimizing apps' quality. They mine keywords and popular topics and conduct sentiment analysis on app reviews, and based on the mining result, they give suggestions on app name, app description, app icon to improve an app's visibility in a certain category. However, they merely help modify the appearance of an app. Therefore, they cannot help users find high-quality apps.

	To address the cold-start problem and recommend high quality apps to users, a fundamental requirement is the capability to accurately measure apps' quality based on the \textit{inborn} features instead of the user-generated information, as a newly published app has very few or even no user's feedback. We face great challenges: (1) \textbf{what inborn features of an app best reflect its quality are unclear.} Thus we performed term frequency counting on 18 million app reviews covering all categories to understand the criteria based on which users rate the apps. The results show that users deeply care about the design of \textit{user interface} (UI), the built-in \textit{advertisements} and requested \textit{permissions} when commenting the quality of an app; (2) \textit{visual design} plays an important role in providing pleasant user experience and \textit{view switching} supports the implementations of different modules' functions in an app, however, \textbf{it is difficult to efficiently extract large amounts of views.} To address the problem, we develop several modules that can automatically record the views, then click a button to switch to the next view and finally traverse all views of an app. We have achieved the goal of analyzing a large number of apps automatically; (3) \textbf {how to precisely measure the importance of the primary and secondary views is a challenge.} Thus we employ a view graph to describe the quality of views and switching among views, and then propose a graph-to-vector method to encode different sizes of view graphs into a sequence of key vertices and their neighbors. Finally, the apps are classified into different quality categories. We aim at enabling the app recommendation system to have a warm-start for those new published apps and those with little user interactions. To highlight, our contributions are: 
	\begin{itemize}
		\item We propose AppQ that leverages apps' inborn features extracted from their source code to grade apps, rather than using extrinsic features like user-generated information, in the aim to warm-starting app recommendation. Static and dynamic analyses are performed in parallel to extract app-level features, which profile apps' behaviors and view-level UI hierarchy as well as the switching among views. To the best of our knowledge, this is the first work on evaluating apps' quality with their inborn features.
		
		\item  We construct an attributed view graph for each app based on view switching. We then encode the graph by a sequence of key vertices and their neighbors to highlight the importance of main views within an app. The graph that represents an app is further converted into a feature vector. 
		
		\item	We employ an ensemble approach on the feature vectors to  classify apps into three quality categories, for ranking out the top-quality ones. With a dataset from Google Play, AppQ achieves the best performance with the accuracy of 85.0\%. The results demonstrate the great promise of warm-starting app recommendation with AppQ.   
	\end{itemize}
	
	The rest of the article is organized as follows. In Section 2, we will introduce some preliminary concepts of android views and review related work. In Section 3, we first introduce the overall framework of the proposed method. Then, we discuss the five procedures in detail, including features identification, feature extraction, view graph construction, graph-to-vector and classification. The evaluation metrics and experimental analysis are demonstrated in Section 4. We will conclude our work in Section 5 with prospective work.

	\section{Background and Related Work}
	
	\subsection{Android Views}
	
\begin{figure}[htb]
		\centering
		\includegraphics[width=.47\textwidth]{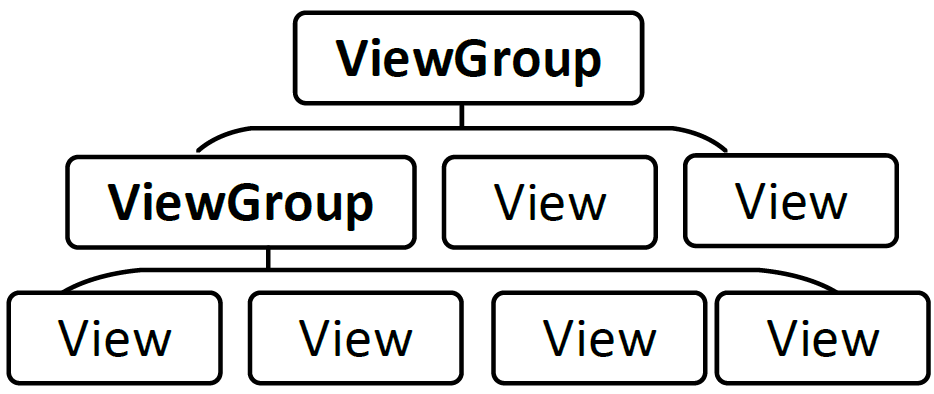}
		\caption{ Illustration of a view hierarchy}\label{layout}
	\end{figure}

\begin{figure}[htb]
		\centering
		\includegraphics[width=.47\textwidth]{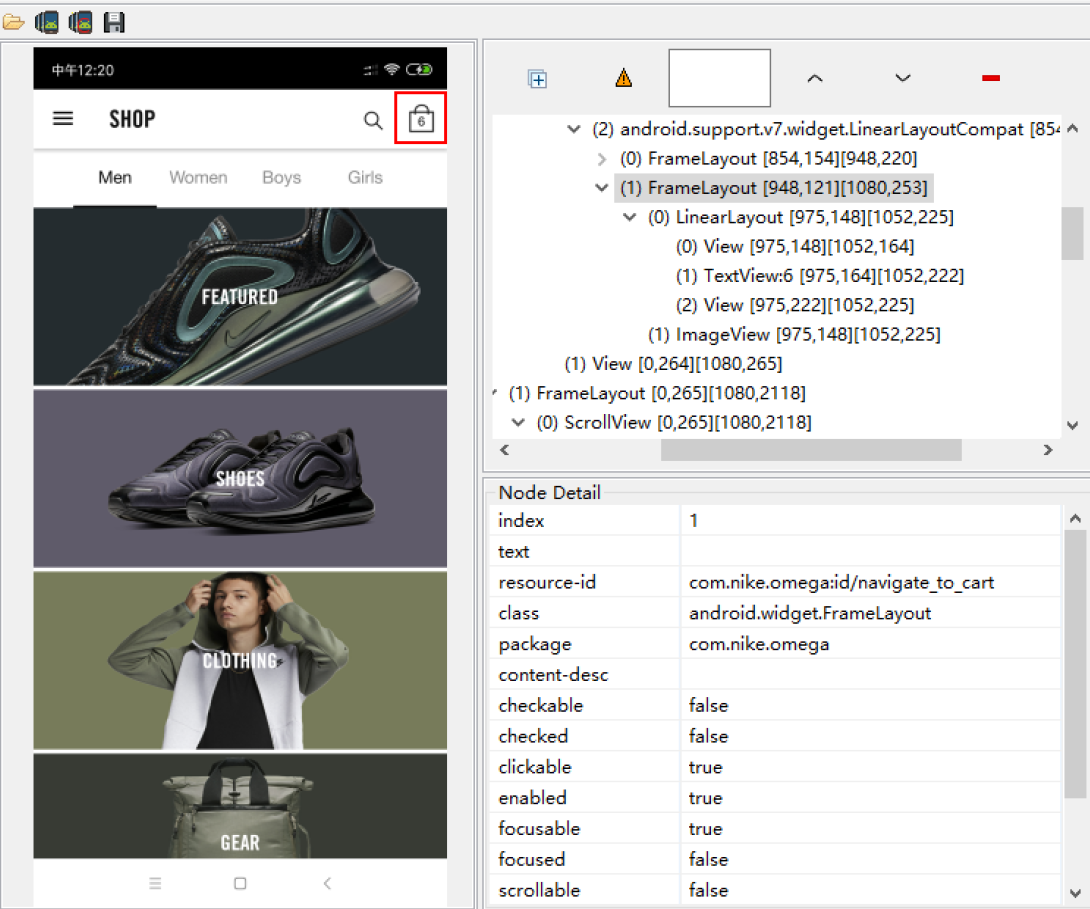}
		\caption{An example of a view and its UI components}\label{fview}
	\end{figure}

	A layout defines the structure for a user interface in the app. All elements in the layout are built using a hierarchy of View and ViewGroup objects. The basic building block for user interface is View which occupies a rectangular area on the screen and is responsible for drawing and event handling. A View provides a screen which users can actually see and interact with \cite{view}. View is the base class for widgets which are used to create interactive UI components like buttons, checkbox, text fields, etc. Whereas a ViewGroup is an invisible container that UI components can be placed in, as shown in Fig.~\ref{layout}. Each subclass of the ViewGroup provides a unique way to display the views. Common layouts include linear layout, relative layout and web view. When users press the buttons on the screen, they are actually interacting with widgets of the view. Uiautomatorviewer is a tool in Android SDK. It provides the information of UI components currently displayed on an Android device. We can inspect the layout hierarchy and view the properties of UI components that are visible on the foreground of the device. For example, in Fig.~\ref{fview}, on the left is a shopping app's view and its UI components are shown on the right. Images and texts are placed in different layouts that are stacked to form the current view. We can see the properties of the red box area from the lower right table, indicating it is focusable and clickable. When users click the red box area, the current view will switch to the shopping cart view. In this paper, we mainly focus on the layout hierarchy of each view and the switching among views.

	\subsection{Related Work}
	Several approaches on recommending and ranking apps from different aspects have been presented in the literature. We conduct a detailed survey of existing work.
	
	Based on user-generated statistics, Pai \textit{et al.} \cite{PaiLWWC17} implemented recommendation by calculating the positive or negative score of semantic orientation in reviews. They also considered statistical information, e.g., star ratings and downloads. Rustgi \textit{et al.} \cite{RustgiFRM17} took four metrics into consideration: downloads, ratings, descriptions and requested permissions. Users are allowed to select the weight of each metric. Then a normalized score is computed for each app. The app with highest scores will be recommended. Jisha \textit{et al.} \cite{JishaKV18} evaluated and clustered apps based on permissions and user ratings. They calculated a risk score and then applied $k$-means to categorize apps into high or low security applications. 
	
	In terms of security, Peng \textit{et al.}~\cite{PengZSHWT18} proposed a matrix factorization algorithm based on permissions and functionalities. They exploited the relationship between the permissions and functionalities to achieve personalized recommendation. Similarly, Zhu \textit{et al.}~\cite{ZhuXGC14} recommended apps by striking a balance between the apps’ popularity and the users’ security concerns, and built an app hash tree to efficiently recommend apps. Liu et at.~\cite{LiuKCGJX15} incorporated interest-functionality interactions and users’ privacy preferences to perform personalized app recommendations. They constructed a model to capture the trade-off between functionality and user privacy preference. Yin \textit{et al.} \cite{YinWCDNH18} considered user interests and category-aware user privacy preferences. They exploited textual and visual content associated with apps to learn multi-view topics for user interest modeling.
	
	Based on users' browsing behaviors and preferences, Shu \textit{et al.} \cite{ShuWLTC018} observed that users first view the description of the app and then decide if they want to download it or not. Thus they proposed ActionRank which integrated various signals from user actions into a coherent model for personalized app recommendations. Similarly, He \textit{et al.} \cite{DBLP:journals/misq/HeFLL19} introduced a model that appropriately combined download and browsing behaviors to learn users’ overall interests of apps. They combined users’ overall interests and their current interests to recommend apps. Xu \textit{et al.} \cite{XuZSY19} observed that a user has a pattern of app usage contexts. What's more, the similarity in two users’ preferences is correlated with the similarity in their app usage context patterns. They proposed a neural approach which learns the embeddings of both users and apps and then predicted a user's preference for a given app.
	
	To make use of the textual information, Pan \textit{et al.} \cite{PanZWX16} proposed a combineLDA method to mine topics of apps based on their descriptions and users' reviews. They calculated the similarity between apps and recommended users with highly similar apps. To increase the usability of emergency apps, Ahmadi \textit{et al.} \cite{DBLP:conf/sigsoft/AhmadiMR19} proposed REMAC  that combined different machine learning techniques to analyze the context characteristics of different organizations and suggested unique features that can be included in the emergency apps. Lin \textit{et al.}~\cite{LinSKC14} noticed that new versions of apps may attract users’ interests, even the apps used to be unappealing. To allow previously disfavored apps to be recommended, they constructed a representation of an app’s version as a set of latent topics from version metadata and textual descriptions. Then they discriminated the topics based on genre information and weighted them on a per-user basis to generate a version-sensitive ranked list of apps for a target user. Liang \textit{et al.} \cite{LiangHLCYW17} considered the interactions among the context information of apps.  They utilized a tensor-based framework to integrate app category information and multi-view features on users and apps. Cao \textit{et al.} \cite{CaoHNWHWC17} integrated both numerical ratings and textual content from multiple platforms. They represented an app as an aggregation of common features across platforms (apps' functionalities) and specific features that are dependent on the resided platform.
	
	Some work focused on other espects, e.g., Su \textit{et al.} \cite{SuLLT17} measured network traffic cost of each app based on random walk. Then they ranked the apps by app rating and traffic cost. Lin \textit{et al.} \cite{LinSKC13} estimated the probability of the user liking the app by extracting information from an app's Twitter followers.
	
	However, most previous related work relies on user-generated statistics, users' browsing behaviors or apps' description contexts, which are scarce for new published apps. Moreover, \cite{Crowdsourced-App-Review-Manipulation-sigir2017} has mentioned that users' reviews to apps are sometimes not reliable. Thus it is necessary to explore the inborn features of apps' view, which are most relevant to user experience, and can be an important indicator of the quality of apps. There are also works on detecting repackaged apps based on UI similarity~  \cite{ZhangHZW014,YueSMT0018,MaoBMJLJ18}, but similar techniques have not been applied in evaluating and ranking apps. In this work, we extract fine-grained features which combined app-level and view-level features to thoroughly profile apps' quality and grade apps.

	\begin{figure*}[htb]
		\centering
		\includegraphics[width=0.85\textwidth]{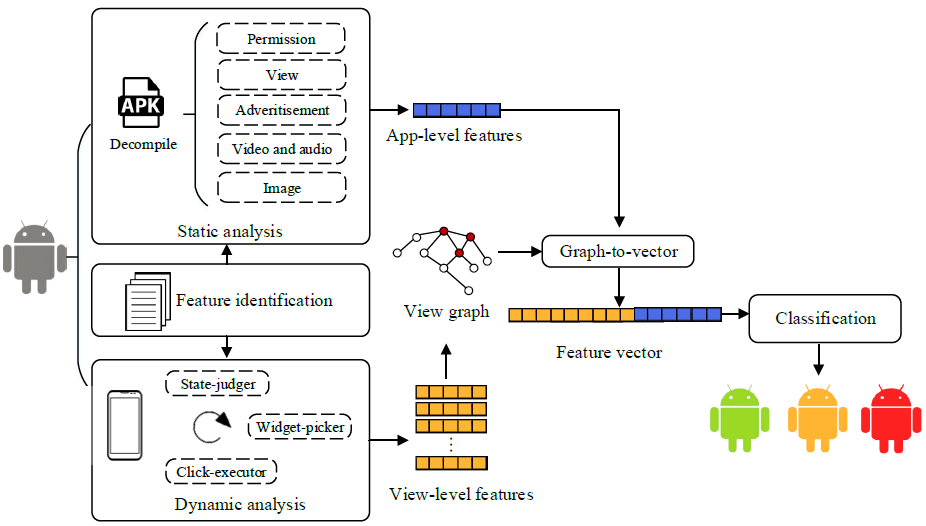}
		\caption{System overview}\label{f_overview}
	\end{figure*}

	\section{Methodology}
	Most current recommendation systems face the cold-start problem because of the applied ranking strategies. To address this problem, we are motivated to recommend apps based on inborn features which can represent apps' real quality, instead of statistical information generated by users. The first question arises: \textit{``what are the inborn features of an app that can reflect its quality?''} To answer this question, we need to understand the criteria on which users based when they rate. We perform term frequency counting on 18 million reviews on Google Play. The results (in Section~\ref{inborn}) show that UI has an important effect on user experience. % Some other concerns are detailed in subsection A. 
	Accordingly, AppQ will focus on modeling app UI features for app quality estimation. More specifically, app-level features and view-level features are considered. App-level features model the higher level features, such as permissions, ads and whether the app crushes during running; view-level features are those related to the design of individual views in an app, such as the widgets enclosed in a view and the layout of a view.  
	%and in parallel, AppQ installs apps and runs them automatically to extract .
	%decompiles apps and extracts app-level features from AndroidManifest.xml and other files. In parallel, AppQ installs apps and runs them automatically. Clickable widgets are triggered to switch views and the layout hierarchy is recorded. The layout of each view makes up the view-level features. The widgets and layouts in one view can reflect apps' UI design, but it is not enough to describe the apps' functions. Apps' functions are implemented by the switching among views. 
	With the features, we will construct an attributed view graph and further converts the graph to %a sequence of key vertices and their neighbors. The combination of view graph and app-level features 
	form the feature vector of the app. Finally, apps will be classified by machine learning algorithms with feature vectors as inputs. The system overview is shown in Fig. \ref{f_overview}
	
	\subsection{Inborn features identification} 
	\label{inborn}
	When rating an app, an individual user may only care certain aspects of an app, so (s)he may only mention a few specific issues in the review. Nevertheless, combining all reviews from all users would give us a complete picture of what truly matters.    
	Hence, we start with analyzing 18 million app reviews covering all app categories from Google Play. 
	
	Specifically, we preprocess the raw data by breaking texts into tokens based on the scikit-learn library \cite{sklearn}), removing the stop words (based on terrier stop words library \cite{stop-word}), removing noise (ill-spelling words) and feature-irrelevant words (e.g., ``app'', ``game'',). After that, we obtain the top-500 high-frequency words related to app features. Table~\ref{term frequency} shows that when users rate apps, they care about the performance of apps at runtime(e.g., crash), the design of user interface (picture and color), the widgets used in an interface (e.g., button), permissions an app may request (e.g., wifi and camera) and advertisements contained in an app (e.g., ads). We divide the inborn features into two groups, app-level features and view-level features. 
	\begin{table}
		% \centering
		\caption{Terms indicating app features in top 500 high-frequency terms}
		{\footnotesize
			\begin{tabular}{p{25pt}p{30pt}p{20pt}p{30pt}p{35pt}p{30pt}} % 6 columns
				\toprule 
				Term & Frequency & Term & Frequency & Term & Frequency \\
				\midrule
				crashes & 424664 & bugs & 99249 & figure & 66207 \\
				graphics & 297585 & photos & 97168 & camera & 62514 \\
				screen & 280734 & access & 96990 & tap & 62505 \\
				features & 178181 & text & 90920 & design & 62220\\
				crashing & 163609 & word & 89965 & internet & 59336  \\
				feature & 157911 & view & 85423 & connection & 56411 \\
				ads & 133841 & pics & 83931 & error & 55946 \\
				crash & 132612 & picture & 81209 & connect & 53777 \\
				touch & 130092 & wifi & 75958 &  color & 52650 \\
				pictures & 126972 & click & 75837 & battery & 50480 \\
				button & 118851 & photo & 75056 & speed & 47885 \\
				interface & 117689 & bug & 73533 \\ 
				\bottomrule
			\end{tabular}
			\label{term frequency}
		}
	\end{table}

	\begin{table}[htbp]
		\caption{Features}
		\begin{center}
			\begin{tabular}{p{3pt}p{141pt}p{30pt}p{30pt}}
				
				\toprule  
				&Features&Dimension&Type\\
				\midrule  
				1&Requested permissions&128&\multirow{6}*{App-level}\\
				2&Number of permissions&1\\
				3&Number of views&1\\
				4&Number of ad libraries&1\\
				5&Number of video or audio files&1\\
				6&Number of images&1\\
				7&Whether the app crashes at runtime&1\\
				\midrule 
				8&Total number of layouts or widgets&1&\multirow{5}*{View-level}\\
				9&Number of different layouts or widgets&1\\
				10&Total depth of layout in current view&1\\
				11&Average depth of layout in current view&1\\
				12&Number of clickable widgets&1\\
				\bottomrule 
			\end{tabular}
			\label{t1}
		\end{center}
	\end{table}

	\subsection{Feature extraction}
	Representative and comprehensive features are in great demand to profile apps' behaviors. We extract app-level features and view-level features by both static analysis and dynamic analysis. The features are summarized in Table \ref{t1}.
	
	\noindent(1) App-level features
	
	AppQ decompiles the APK to access the inner files. Android controls access to system resources with permissions. Specific permissions are required when an app interacts with system APIs or databases. To characterize an app's intention during resource accessing, we extract the requested permissions from its manifest file. Only system-defined permissions are considered. Besides, all displayed views are also registered in the manifest file, which we can make use of. The number of declared views reflects the opportunity of interactions provided by apps. Usually the more views there are, the more functions the app has. In addition, the number of multimedia files in the payload mirrors the aesthetics of apps. We count the number of video(filename extension as ``.mp4'',``.avi'',``.wmv'' and ``.flv''), audio(filename extension as ``.mp3'') and images(filename extension as ``.jpg'',``.jpeg'',``.png'',``.bmp'' and ``.gif'') in the APK as features.	Advertisement has a negative impact on user experience. We collect 67 popular advertisement libraries, e.g., Google Ads and Amazon Ads, and take the number of libraries in the APK as a feature. Whether the app crashes at runtime shows how the app performs in delivering what it has promised, which is one of users' most concerned issues. Most Android-based mobile systems (e.g., MIUI) can pop up an alarm window to remind users the app has stopped running  when a crash occurs. AppQ records the alarm as a feature. Finally, a 134-dimensional app-level feature vector is formed.
	
	\noindent(2) View-level Features
	
	Since static code analysis faces the challenges raised by code obfuscation and dynamic code loading, we dynamically run each app to extract the hierarchy of each view and the switching among views. 
	
	The number of layouts or widgets reflects the layering of the view. The number of different layouts or widgets represents the diversity of widgets of in the view. The widgets can be set as ``clickable’’ thus users can click them to trigger some events, e.g., login or jump to the next view. We take the number of clickable widgets as a feature to describe the opportunities the app provides to interact with users. These parameters of a complex interface with a layered view are larger than those of a simple view. The features also reflect the developers’ proficiency in UI development. Finally, we transform each view's hierarchy into a 5-dimensional feature vector.
	
	AppQ can automatically install, run and uninstall apps with the help of AndroidViewClient \cite{AndroidViewClient} and adb commands \cite{Adb}. AppQ is technically novel in extracting apps' views. When running the app, AppQ takes depth-first search to traverse clickable widgets. There are three alternately working modules: state-judger, widget-picker and click-executor. 
	
	\noindent \textbf{State-judger: }State-judger is to judge whether the view has changed after each click and then performs corresponding operations according to the judgment. It makes use of AndroidViewClient to dump the hierarchy of the current view, including the information of layouts and widgets, e.g., package name, class name. A problem arises that AndroidActivityClient does not provide information about the activity id. Each activity needs a unique id so that AppQ can recognize it when jump to the same activity. To deal with the problem, AppQ crawls the hierarchy of the activity and collects class names of each layout or widget which are further concatenated into a string. The hash value of the string is regarded as the activity’s id which is the flag of judging whether the activity has changed. Since activities can start other activities which live in a separate app, AppQ also needs to determine whether current activity belongs to the test app. If not, it triggers back button to return to previous activity. Consequently, AppQ is able to take the corresponding operations according to the state of a view: (a) If a new view is on display, AppQ records the switch and dumps its hierarchy; (b) If the view has not changed and there exist clickable widgets in it remaining unclicked, AppQ chooses the one with the highest priority from the waiting list. The priority is set by the widget-picker module; (c) If the view has not changed and all clickable widgets in it have been clicked, AppQ triggers the \textit{back} button to return to the previous view.
	
	\vspace{10pt}
	\noindent \textbf{Widget-picker: }Widget-picker is to determine the clicking sequence of widgets. Different types of widgets have different functions, e.g., EditText is for entering and modifying text while ImageButton displays a button with an image (instead of text) that can be clicked. To improve the efficiency of triggering the switching among views, we set the priority of widgets based on their locations and types. Firstly, AppQ extracts clickable widgets in the current view. Though some widgets are clickable, they are unable to trigger view switches, e.g., ToggleButton, CheckBox, RatingBar, etc. Thus they are filtered out. Secondly, the priorities are determined by widgets' positions. Upper left corner is normally the return button or the navigation drawer containing app setting which we are less interested in. Thirdly, for widgets in the same area, their priorities follow ImageButton\textgreater Button\textgreater ImageView\textgreater TextView\textgreater other widgets. Our testing approach is more efficient than randomly clicking tools such as Monkey \cite{Monkey}.
	
	\noindent \textbf{Click-executor: }Click-executor is to execute click commands. State-judger sends click command to click-executor to click a widget or return to the previous view.
	
	The total number of layouts reflects the

	\subsection{View graph construction}
	A well-defined graph reveals relationship between an app and its views, not only by features, but also by view neighbor relationship. An app implements different functions through view switching. After feature extraction, we have obtained feature vectors that profile apps' behaviors and the hierarchy of views. We build an undirected attributed view graph for each app, which is defined as $G(V,E,F)$. $V$ is the vertex set and each $v \in V$ represents a view.  $E$ is a set of edges. $(v_1,v_2)$ means $v_1$ can switch to $v_2$ by clicking some widgets. Technically each view is able to return to previous view by clicking return button. $F$ is the feature set that contains only view-level features extracted from dynamic analysis. App-level features will be merged during Graph-to-vector step introduced below.
	
	\subsection{Graph-to-vector}
	After feature extraction and graph construction, we have obtained app-level features and attributed view graphs. To classify apps into different quality categories, the following problems need to be addressed: 
	
	a. View graphs differ in size for different apps, how to transform them into a standardized expression? 
	
	b. How to merge app-level features and the view graph  to profile one app?
	
	Inspired  by the recent  explosion of graph neural network (GNN) models, such as GCN \cite{KipfW17}, GraphSAGE \cite{GraphSage}, GAT \cite{GAT} and GIN \cite{GIN19} that can be used to learn a representation vector for every graph, the most feasible solution for us is to map each app as a vector by integrating the app-level features and the view graph. GNN models are usually defined as multi-layer neural networks with layer-wise propagation operators on graph datasets. The main intuition is to fully utilize both the features of the nodes and the structural information of the graph in node/graph classification or label prediction problems. GCN \cite{KipfW17} is derived from a first-order approximation of graph Laplacian spectrum. The normalized adjacency matrix serves as a new convolution kernel of traditional CNN which can incorporate additional information from correlative node items. In this way, one graph convolutional layer can be viewed as gathering features from 1-hop neighbours of the central node. By stacking multiple convolutional layers, we can capture further information through the graph structure.
	%The convolution layer is expressed as:}
	%\begin{equation}
	%   H^{(l+1)}=\delta(\hat{D}^{-1/2})\hat{A}\hat{D}^{-1/2}H^{(l)W^{(l)}}
	%\end{equation}
	%\textcolor{red}{where $\hat{A}=A+I_N$ is the adjacency matrix with self-loops. $\hat{D}_{ii} = \sum_j\hat{A}_{ij}$ is the diagonal degree matrix. $W$ is the $l$-th layer weight matrix, and $\delta(\cdot)$ is the activation function. $H^{(l)}$ is the input features of nodes in layer $l$. $\delta$ is a nonlinear activation function. }
	
	Instead of directly using graph Laplacian operator to aggregate information from nodes' neighbours, GAT \cite{GAT} tries to differentiate the contributions of various features of neighbouring nodes to the central node. They first apply a shared self-attention mechanism on the input features to get the attention coefficients, then the coefficients are passed to a softmax function to get the normalized attention weights.
	%\begin{equation}
	%    \alpha_{ij}=
	%    \frac{exp(LeakyReLU(\vec{a}^T[W\vec{h_i}||W\vec{h_j}]))}{\sum_{k\in{N_i}}exp(LeakyReLU())}
	%\end{equation}
	%\textcolor{red}{where $N_i$ denotes the neighbours of node $i$. $||$ is the concatenation operation. $W$ is the shared weight matrix.}
	Then the normalized attention weights are used to aggregate transformed features from neighbouring nodes.
	%\begin{equation}
	%    \hat{h}'_i=\delta \left(\sum_{j\in N_i}\alpha_{ij} W\vec{h_j}\right)
	%\end{equation}
	To better learn the hidden features in different sub-spaces, usually the multi-head attention mechanism is employed on the same inputs.
	%\begin{equation}
	%    \hat{h}'_i=\mathop{\parallel}\limits ^{K}_{k=1}\delta \left(\sum_{j\in N_i}\alpha_{ij} W\vec{h_j}\right)
	%\end{equation}
	%\textcolor{red}{where $K$ denotes the number of attention heads and $\parallel$ denotes the concatenation of features from $K$ heads.}
	
\begin{figure}[htb]
		\centering
		\includegraphics[width=.5\textwidth]{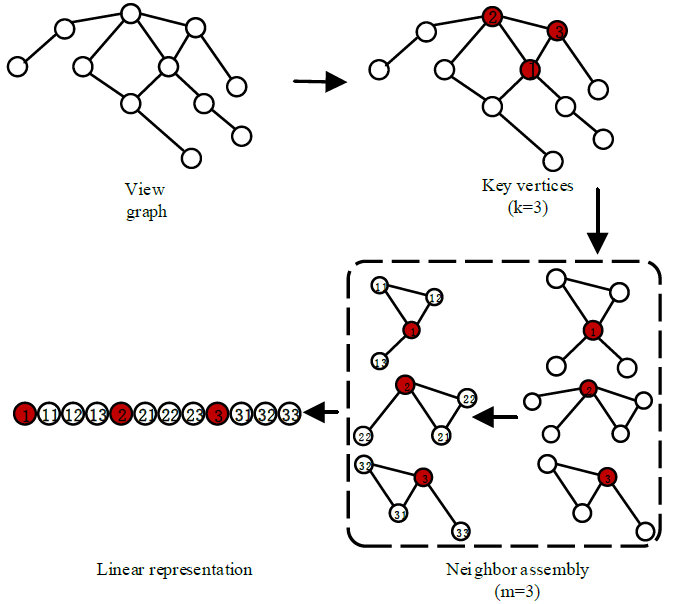}
		\caption{Graph-to-vector method}\label{f4}
	\end{figure}

	Although usable, the recent GNN-based models cannot appropriately handle the special graphs in our case. The main reason is that our vertices (views) have special dependencies and significances in the neighborhood of different vertices when evaluating the quality of an app. Considering those insignificant vertices (views) will introduce too much noise to the model. %while GNN models apply the same weights when aggregating the neighborhood for each vertex.
	But GNN models apply the same aggregating operators to all the neighborhood for each vertex. Thus, we focus on the key vertices and their neighbors when representing an app as a vector, rather than applying the learned aggregator iteratively  on each vertex's neighborhood. Experimental comparison also shows that our   next presented graph-to-vector conversion approach performs better than the recent GNN models. Specificlly, we apply the following steps to each graph: (1) select a fixed number of key vertices from the graph; (2) assemble a fixed-size neighborhood for each key vertex; (3) construct feature vectors. The process is shown in Fig. \ref{f4}.

	\vspace{8pt}
	\noindent(1) Key vertices selection
	
	Since the main views of an app are  the most important ones regarding the quality of the app, we measure the importance of vertices within a graph by their centrality: 
	\begin{equation}
	d_i={\sum}_{i \in V, j \in V, i \neq j}distance(v_i,v_j)\label{eq1}
	\end{equation}
	where $distance(v_i,v_j)$ is the shortest distance between vertex $v_i$ and $v_j$. The smaller $d_i$ is, the higher  centrality of $v_i$ will be, and   $v_i$ locates at a more central position in the graph. In the view graph, some views can switch to many other views, whereas some have no branches at all. The ones with more switches are more centralized, indicating that they are more important in the graph structure. We calculate the centrality of each vertex and select the top $k$ vertices as key vertices. The steps are described in Algorithm 1.
	
	\begin{table}[htbp]
		\begin{center}
			\begin{tabular}{l}
				\toprule  
				\textbf{Algorithm 1:  Key vertices selection}\\
				\midrule  
				\textbf{Input:} $G(V,E)$, $k$\\
				\textbf{Output:} Key vertices $L_k$\\
				1. \textbf{for} $v_i \in V$\\
				2. \quad calculate $d_i=\sum_{i \neq j}Dijkstra(v_i,v_j)$,\\
				3. \quad add $d_i$ in centrality list $L_d$\\
				4. \textbf{end for}\\
				5. 	$sort( L_d,reverse=False)$\\
				6. \textbf{if} $k \leq L_d$\\
				7. \quad select top $k$ vertices $L_k=[v_1,\dots,v_k]$ form $L_d$\\
				8. \textbf{else}\\
				9. \quad add $k-|L_d |$ dummy vertices to $L_k$\\
				10. \textbf{return} key vertices $L_k$\\
				\bottomrule 
			\end{tabular}
			\label{Algorithm1}
		\end{center}
	\end{table}

	\noindent(2) Neighbor assembly
	
	For each key vertex $v$ from the previous step, $m$ neighbors are supposed to be extracted. We sort $v's$ 1-hop-neighbours, 2-hop-neighbours and 3-hop-neighbours by centrality and regarded them as candidate vertices. Since different $v$ may have a different number of candidate vertices, we only pick top $m$ of them. If the number of candidate vertices is smaller than $m$, the algorithm creates dummy vertices with all-zero features for padding purposes. Now the arbitrary graph can be represented by $k*(m+1)$  vertices:
	\begin{equation}
	\begin{split}
	[v_1,\underbrace{n_{11}^1,\dots,n_{1a}^1,n_{21}^1,\dots,n_{2b}^1,n_{31}^1,\dots,n_{3c}^1}_{m}]\dots\\
	[v_k,\underbrace{n_{11}^k,\dots,n_{1a'}^k,n_{21}^k,\dots,n_{2b'}^k,n_{31}^k,\dots,n_{3c'}^k}_{m}]\label{eq2}
	\end{split}
	\end{equation}
	where $v_k$ is a key vertex, $n_{3c^{'}}^k$ means the $c^{'}$-th 3-hop-neighbor vertex of $v_k$. For vertices with higher centrality, they may appear as neighbors of several vertices, indicating they have higher impact to the graph. The priority can be reflected in the feature vector. The steps are described in Algorithm 2.

	\begin{table}[htbp]
		\begin{center}
			\begin{tabular}{l}
				\toprule  
				\textbf{Algorithm 2:  Neighbor assembly}\\
				\midrule  
				\textbf{Input:}  $G(V,E)$, key vertices $L_k$, centrality list $L_d$, $m$\\
				\textbf{Output:} Assembly list $L_a$\\
				1. \textbf{for} $v_i \in L_k$\\
				2. \quad get $v_i's$ 1-hop-neighbors $N_1^i=[n_{11}^i,n_{12}^i…n_{1a}^i]$\\
				3. \quad get $v_i's$ 2-hop-neighbors $N_2^i=[n_{21}^i,n_{22}^i…n_{2b}^i]$\\  
				4. \quad get $v_i's$ 3-hop-neighbors $N_3^i=[n_{31}^i,n_{32}^i…n_{3c}^i]$\\
				5. \quad $sort(N_1^i,reverse=False)$ according to $L_d$\\
				6. \quad add $N_1^i$ in candidate list $L_c$\\
				7. \quad \textbf{if} $|L_c |<m$\\
				8. \quad \quad $sort(N_2^i,reverse=False)$ according to $L_d$\\
				9. \quad \quad add $N_2^i$ in candidate list $L_c$\\
				10. \quad \textbf{if} $|L_c |<m$\\
				11. \quad \quad $sort(N_3^i,reverse=False)$ according to $L_d$\\
				12.  \quad \quad add $N_3^i$ in candidate list $L_c$\\
				13. \quad \textbf{if}  $|L_c |<m$\\
				14. \quad \quad add dummy vertices to $L_c$\\
				15. \quad select top $m$ vertices $[n_1^i,\dots,n_m^i]$ from $L_c$\\
				16.  \quad add $v_i$ in assembly list $L_a$\\
				17.  \quad add $[n_1^i,\dots,n_m^i ]$ in assembly list $L_a$\\
				18. \textbf{end for}\\
				19. \textbf{return} assembly list $L_a$\\
				\bottomrule 
			\end{tabular}
			\label{Algorithm2}
		\end{center}
	\end{table}

	\noindent(3) Feature vector construction
	
	Note that each of the $k*(m+1)$  representative vertices has its view-level features. To form a feature vector for an app, we arrange these vertices' view-level features in order and add app-level features at the end.
	
	The app-level features are denoted as
	\begin{equation}
	F_{app}=[f_1,\dots,f_n ]\label{eq3}
	\end{equation}
	and each view's view-level features are
	\begin{equation}
	F_{view}^i=[f_1^i,\dots,f_{n'}^i ]\label{eq4}
	\end{equation}
	where $n$ and $n'$ are the numbers of features. Views in one app may have different view-level features, but they have the same app-level features. Based on \eqref{eq2}, $F(G)$ encodes both app-level features and view-level features, which can be expressed as 
	\begin{equation}
	\begin{split}
	F(G)=[F_{view}^{v_1},\underbrace{F_{view}^{n_{11}^1},\dots,F_{view}^{n_{3c}^1}}_{m},\dots,\\
	F_{view}^{v_k},\underbrace{F_{view}^{n_{11}^k},\dots,F_{view}^{n_{3c'}^k}}_{m},F_{app}]\label{eq5}
	\end{split}
	\end{equation}
	Then a unique mapping from a graph representation into a vector space representation is formed. 
	This feature construction is analogous to the concatenation aggregator used in GNN-based models \cite{GraphSage} and \cite{NiepertAK16}. In the next experimental evaluation section, we will demonstrate that our selected representative vertices by node centrality well serves our study purpose on   characterizing apps' quality for user experience. The feature vector is further fed into the classification algorithms for quality evaluation. The steps are described in Algorithm 3.
	
	\begin{table}[htbp]
		\begin{center}
			\begin{tabular}{l}
				\toprule  
				\textbf{Algorithm 3: Feature vector construction}\\
				\midrule  
				\textbf{Input:} assembly list $L_a$, $F_{view}$, $F_{app}$\\
				\textbf{Output:} feature vector $F(G)$\\
				%1. Initialize vector $F(G)$ to $\vec{0}$\\
				1. \textbf{for} $v_i \in L_a$\\
				2. \quad add $F_{view}^i$ to $F(G)$\\
				3. \textbf{end for}\\
				4. add $F_{app}$ to $F(G)$\\
				5. \textbf{return} $F(G)$\\
				\bottomrule 
			\end{tabular}
			\label{Algorithm3}
		\end{center}
	\end{table}

	\subsection{Classification}\label{AA}
	We employ three machine learning algorithms as classifiers, namely, Linear Support Vector Machine (Linear-SVM), Rbf-kernel Support Vector Machine (Rbf-SVM), and Random Forest (RForest) \cite{sklearn}. To take advantage of the strengths of each classifier and compensate for the disadvantages, we apply ensemble learning to obtain the final detection. These classifiers first work in parallel to predict the label of an app. Then the final label is determined by their majority voting, e.g., ``high quality'' or ``low quality''. The ensemble decision is more stable than single machine learning algorithm.

	\section{Evaluation}
	In this section, we conduct a series of experiments to evaluate AppQ's accuracy and efficiency. We collected 3050 apps from 16 categories in Google Play as our experiment dataset. We also crawled the statistical information, including download numbers, rating scores, number of ratings and number of reviews for assigning apps' labels. The apps have very different statistical information, raging from 0 to 1 billion. Some apps are popular while some have little users' feedback. The description of apps is shown in Table \ref{t2}.

	In each category, the apps are divided into three quality groups. For each group, 75\% apps are used for training and 25\% for testing. We use $n$-fold cross-validation in our experiments to eliminate contingency.

	The experiments are constructed on a computer with two quad-core 2.6 GHz i5 processors and 8G memory. During dynamic analysis, apps are run for 3 minutes on Xiaomi Mi 8 with MIUI based on Android 8.1. AppQ is developed in Python 2.7.

	\begin{table*}[htbp]
		\caption{Description of apps in out dataset. The apps have very different statistical information, raging from 0 to 1 billion. Some apps are popular while some have little users' feedback.}
		\begin{center}
			\begin{tabular}{lcllllllll}
				\toprule  
				\multirow{2}*{Category}&Number of&\multicolumn{2}{l}{Downloads}&\multicolumn{2}{c}{Rating}&\multicolumn{2}{c}{Number of ratings}&\multicolumn{2}{c}{Number of reviews}\\
				
				\quad&samples&Min&Max&Min&Max&Min&Max&Min&Max\\
				\midrule
				Vehicles&74&10&10,000,000&0.0&5.0&0&414,955&0&184,064\\
				Books&254&100&1,000,000,000&0.0&5.0&0&3,079,596&0&935,519\\
				Education&145&1000&100,000,000&1.4&5.0&2&6,929,028&0&2,626,518\\
				Events&180&1&5,000,000&0.0&5.0&0&47,209&0&14,217\\
				Food&227&10&50,000,000&0.0&5.0&0&1,112,158&0&419,931\\
				Health&195&500&500,000,000&1.7&5.0&4&4,734,528&1&1,129,616\\
				House&147&10&100,000,000&0.0&5.0&0&725,959&0&276,617\\
				Music&263&1000&1,000,000,000&2.1&5.0&5&13,782,152&1&4,078,292\\
				News&328&100&1,000,000,000&0.0&5.0&0&12,502,733&0&3,080,332\\
				Shopping&226&100&100,000,000&1.0&5.0&1&7,401,417&0&2,665,869\\
				Social &201&1000&1,000,000,000&1.0&5.0&0&83,539,528&0&22,854,161\\
				Sports&193&100&50,000,000&1.0&5.0&1&1,034,292&0&349,972\\
				Tools&136&1000&1,000,000,000&1.0&4.9&0&43,865,755&0&11,160,716\\
				Travel&134&100&1,000,000,000&1.0&5.0&1&9,837,593&0&2,398,866\\
				Weather&175&100&100,000,000&1.0&5.0&0&2,407,160&0&607,089\\
				Photography&173&1000&1,000,000,000&1.1&4.9&8&14,411,700&7&3,593,948\\
				Total&3050\\
				\bottomrule
				
			\end{tabular}
			\label{t2}
		\end{center}
	\end{table*}

	\subsection{Assigning labels} \label{Assigning-labels}
\begin{figure}[htb]
		\centering
		\includegraphics[width=0.49\textwidth]{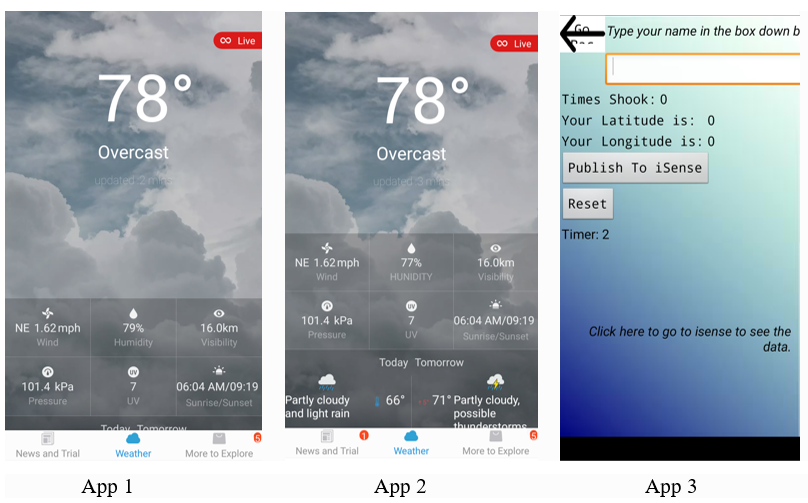}
		\caption{Example apps whose labels are shown in Table \ref{t_label}. Here App 1 and App 2 are similar apps from the same developer. App 3 is a low quality app.} \label{f_label}
	\end{figure}

	Classifying apps into different quality categories needs ground truth labels for training and evaluation. While it may sound intuitive to consider rating as the ground truth for app quality, rating alone is not reliable as it can be easily manipulated~\cite{ChenHZY17}. Moreover, other factors such as number of downloads, number of ratings, and number of reviews are also used by app stores for app ranking.     
	To reduce the risk of ratings being tampered with, 
	we assign labels based on the following two steps: (1) We cluster the apps into three classes by Agglomerative Clustering \cite{sklearn} based on four factors: downloads, ratings, the number of ratings and the number of reviews. 
	%Since the weights of these four factors are unknown, to approximate the app's quality as accurate as possible, we assign three groups of weights to them, which are (a) 0\%, 50\%, 30\%, 20\%. (b) 40\%, 30\%, 20\%, 10\%. (c) 10\%, 40\%, 30\%, 20\%. 
	We apply three groups of labels in this evaluation: 
	1 represents ``low quality'', 2 represents ``average quality'' and 3 represents  ``high quality''. (2) We manually check the labels. We observe that some apps in the same category come from the same developer and have similar functions, UIs and package names. However, the number of downloads and reviews differ greatly. For instance, in category Weather, ``Live Weather \& Local Weather'' has five million downloads while ``Live Weather \& Widget for Android'' has only ten thousand. The developer released them at different times, resulting in more cumulative downloads for the first released apps. Based on step (1), these apps would be clustered into different classes since their four factors could vary widely, hence introducing noise into the training dataset due to the similar UIs. To handle this problem, we collect these similar apps based on their package names and sizes, and unify their labels into the highest label among them. For other ``low quality apps'' from step (1), we manually check them and only keep the ones with badly-designed interfaces. Examples are shown in Fig. \ref{f_label} and Table. \ref{t_label}.
	
	\begin{table}[htbp]
		\caption{Example apps to illustrate the label assignment process. Here App 2's label is corrected due to the step 2 of our labelling process.}
		\begin{center}
			\begin{tabular}{clcc}
				\toprule  
				Index&App name&Step (1)&Step (2)\\
				&&label&label\\
				\midrule  
				1&Live Weather \& Local Weather&3&3\\
				2&Live Weather \& Widget for Android&1&3\\
				3&Shake Alert&1&1\\
				\bottomrule
				\multicolumn{4}{l}{App 1 and App 2 have similarity package names, which are:}\\
				\multicolumn{4}{l}{mobi.infolife.ezweather.widget.storm}\\
				\multicolumn{4}{l}{mobi.infolife.ezweather.widget.raining.glass}\\
			\end{tabular} 
			\label{t_label}
		\end{center}
	\end{table}

	\subsection{Metrics}\label{AA}
	Two metrics are applied to evaluate the classification result of AppQ, Micro-F1 and Dissimilarity (Diss).
	%Accuracy is defined as
	%\begin{equation}
	%Accuracy=\frac{True}{True+False}\label{eq6}
	%\end{equation}
	
	The F1-score is a measure of a test's accuracy and is used to evaluate the performance of unbalanced classification problem, as we have more label-3 apps in some categories of our dataset. However, F1-score is for binary classification problems. We apply Micro-F1 for multilabel classification. It calculates metrics globally by counting the overall true positives, false negatives and false positives. In the following, we regard Micro-F1 as the accuracy measurement:
	\begin{equation}
	Micro-F1=2\frac{recall_{micro}\times precision_{micro}}{recall_{micro}+ precision_{micro}}
	\end{equation}
	\begin{equation}
	precision_{micro}=\frac{\sum_{i=1}^k TP_i}{\sum_{i=1}^k TP_i+FP_i}
	\end{equation}
	\begin{equation}
	recall_{micro}=\frac{\sum_{i=1}^k TP_i}{\sum_{i=1}^k TP_i+FN_i}
	\end{equation}
	where $k$ is the number of label categories and TP, FP, TN, FN denote true positive, false positive, true negative, false negative, respectively. 
	
	We measure the dissimilarity between predicted labels and the ground truth labels by
	\begin{small}
		\begin{equation}
		Dissimilarity=\frac{\sum_{i=1}^n|predicted label(i)-ground truth label(i)|}{the \; number \; of \;  false \; predictions}
		\end{equation}
	\end{small}
	where $n$ is the number of apps in each category.

	\subsection{Classification results}
\begin{figure*}[htb]
		\centering
		\includegraphics[width=0.83\textwidth]{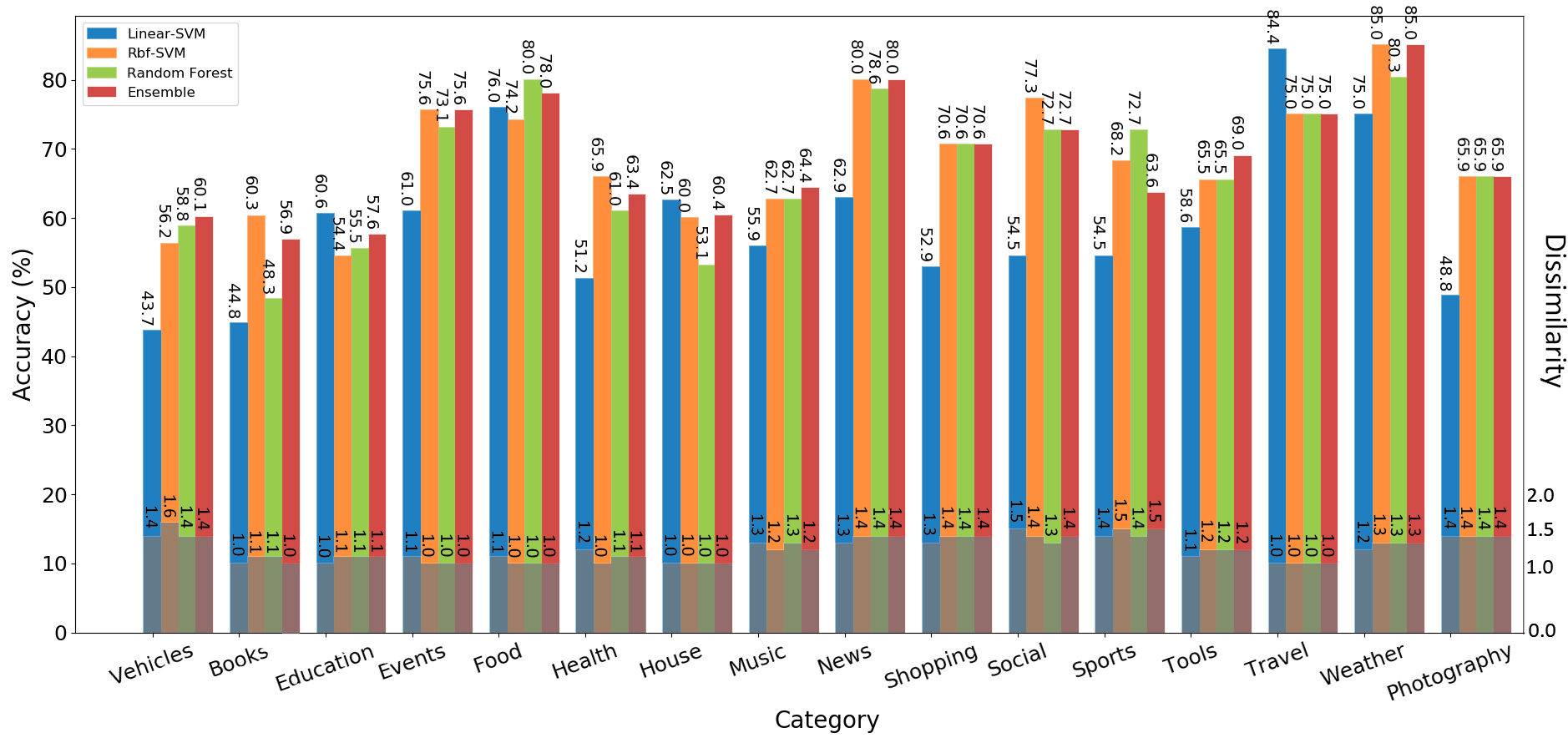}
		\caption{Classification results}\label{f5}
	\end{figure*}

	In this subsection, we evaluate AppQ's accuracy and efficiency. As shown in Fig. \ref{f5}, by employing majority voting mechanism in classification, we achieve the highest accuracy of 85.0\% and dissimilarity of 1.3 on Weather category. It has satisfying performance on the remaining categories
	with accuracy around 60\%-80\%, except for Books and Education. AppQ reaches the lowest accuracy of 56.9\% on Books, which can be attributed to the noisy dataset. Many apps with different functions are placed in the same category in Google Play. For instance, Books category contains online eBooks, file readers and dictionaries, whose view qualities differ greatly. Online eBooks tend to have fancy and attractive views with lots of pictures. In contrast, file readers' designs are normally unobtrusive and simple. Their users only need to select the format of files, e.g., pdf or text, and then the app scans the smartphone and shows the related documents. Differently, the apps in the weather category are more similar because their functions are more concentrated on providing weather forecast. AppQ achieves much better results on categories which contain apps with similar purposes. 
	
	The parameter selection is described as follows:
	First, we use a single classifier on each category with different ground truth labels generated by different weights of the four factors, as described in section \ref{Assigning-labels}. We count the number of times of best average accuracy based on three weight groups on 16 categories. The weight group (0\%, 50\%, 30\%, 20\%) achieves 11 best average accuracy results.  According to our observation, comparatively, some apps have more ratings and reviews even with fewer downloads, as an example shown in \ref{tdownlaods}. Therefore, it is reasonable to assign low weights to downloads. In the following evaluations, we apply the labels generated by weight group (0\%, 50\%, 30\%, 20\%) of the four factors as the ground truth labels. For single classifiers, the average accuracy of Rbf-SVM is the highest, which is 84.2\%. The accuracy of ensemble method is not always the best, but it can balance the results of each classifier so that the overall accuracy is more stable than a single classifier. 
	\begin{table}[htbp]
		\caption{The reason to assign low weight to downloads}
		\begin{center}
			\begin{tabular}{lll}
				\toprule  
				&Generic&Geotracker\\
				\midrule  
				Downloads&50,000,000&1,000,000\\
				Ratings&4.0&4.5\\
				Number of ratings&17,111&47,735\\
				Number of reviews&5,071&16,151\\
				\bottomrule 
				\multicolumn{2}{l}{Generic: com.vznavigator.Generic}\\
				\multicolumn{2}{l}{Geotracker: com.ilyabogdanovich.geotracker}\\
			\end{tabular}
			\label{tdownlaods}
		\end{center}
	\end{table}
	
	The number of vertex we extracted in each app is in range [6, 10]. For the selection of $k$ and $m$ in the graph-to-vector phase, we set $k$ in range [4, 6] and $m$ in range [0, 3]. We conduct several experiments with different combinations of $k$ and $m$. When $k=4$ and $m=2$, it achieves the best performance.

	\subsection{Feature comparison}
\begin{figure}[htb]
		\centering
		\includegraphics[width=0.47\textwidth]{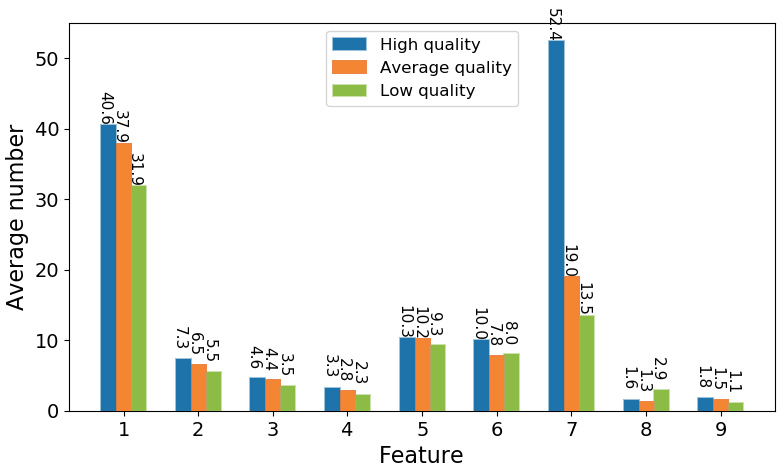}
		\caption{Feature comparison in Weather category
			1: Total number of layouts or widgets; 
			2: Number of different layouts or widgets; 
			3: Total depth;
			4: Average depth; 
			5: Number of clickable widgets; 
			6: Number of permissions; 
			7: Number of views; 
			8: Number of ad libraries; 
			9: Number of video or audio files.
		}
		\label{fweather_f}
	\end{figure}

\begin{figure}[htb]
		\centering
		\includegraphics[width=0.48\textwidth]{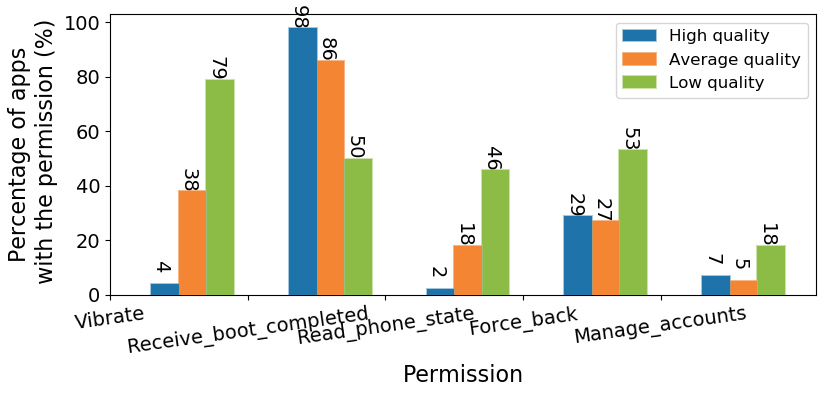}
		\caption{Permission comparison in Weather category}\label{fweather_p}
	\end{figure}

	Behaviors of apps are reflected by features. To figure out the differences between high quality and low quality apps, we take further analysis of their features. Take the Weather category as an example. As shown in Fig.~\ref{fweather_f}, the average numbers of view-level features of high quality apps are higher than those of low quality apps, especially on the number of widgets. It indicates that the layout of high quality apps are more carefully designed, and they are more complicated and layered. For app-level features, high quality apps tend to have more views registered in their manifest files. The more views there are, the more functionalities an app has. Advertisement is another factor users care about. From the figure we can see there are less advertisement libraries in high quality apps. It confirms that  advertisement does have a negative impact on user experience. Due to privacy concerns, users give lower ratings to those with permissions to change the state of smartphones, e.g., vibrate or force back, as shown in Fig. \ref{fweather_p}.

	Some features are related to apps' functions, so there are no common patterns among different categories of apps. For instance, all apps in Weather have few video or audio files despite their quality. It is because most app-level features we extracted are from static analysis, but some videos and audios are dynamically downloaded due to the real-time nature of apps in Weather category. But for apps in categories that do not always require real-time services, e.g., Education, the number of videos in high quality and low quality apps differ greatly, which is 29.4 and 2.1, respectively.
	
	\subsection{Time consumption}
	To evaluate the time consumption of AppQ, we analyze the execution time of each phase, which is shown in Table \ref{t7}. In the app-level feature extraction phase, decompiling is the most time consuming step. On average it takes 17 seconds to decompile an app and takes 1 second to extract app-level features. In the view-level feature extraction phase, AppQ runs each app for 3 minutes. Since the extraction of these two kinds of features work in parallel, the overall time consumption of feature extraction phase depends on the one that takes longer time, which is 183 seconds by view-level features extraction. Graph-to-vector step is very light-weighted, taking 0.003 second on each app. Then on the classification step, classifiers work in parallel. The prediction process does not take much time. Finally, we apply the ensemble results of the classifiers. In general, it takes an average of 183 seconds to analyze each app. AppQ is efficient and adequate for classifying large amounts of apps in the markets, especially when it is applied to analyze new apps.
	
	\begin{table}[htbp]
		\caption{Time consumption (second)}
		\begin{center}
			\begin{tabular}{ccccc}
				\toprule  
				\multicolumn{2}{c}{I}\\
				a&b&II&III&Total\\
				\midrule  
				18&183&0.003&0.005&183\\
				\bottomrule
				\multicolumn{5}{l}{I. Features extraction}\\
				\multicolumn{5}{l}{a. App-level features extraction}\\
				\multicolumn{5}{l}{b. View-level features extraction}\\
				\multicolumn{5}{l}{a and b work in parallel}\\
				\multicolumn{5}{l}{II. Graph-to-vector III. Classification}\\
				
			\end{tabular}
			\label{t7}
		\end{center}
	\end{table}
	
	\subsection{Comparison with related work}
\begin{figure}[htb]
		\centering
		\includegraphics[width=0.48\textwidth]{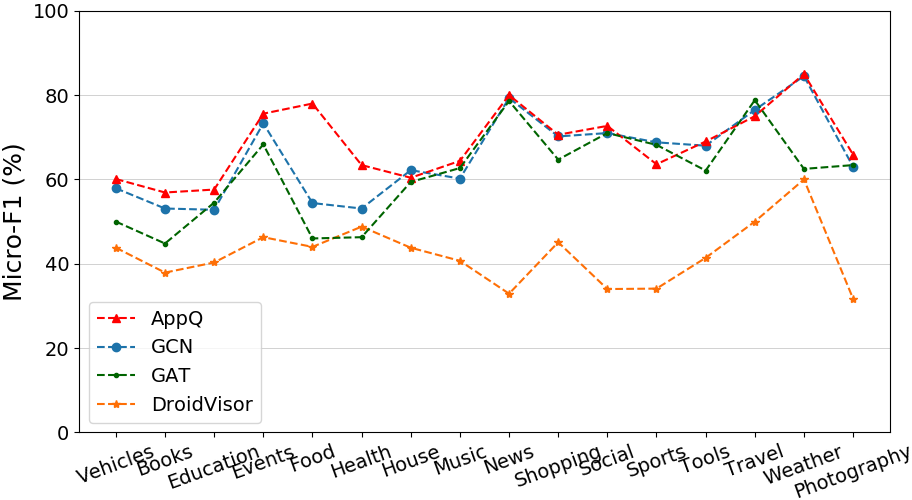}
		\caption{Comparison with DroidVisor, GCN and GAT}\label{compare}
	\end{figure}

	Most of existing work on recommendation is based on natural language processing. They get users' feedback from reviews to recommend potentially interesting apps \cite{FuLLFHS13}.  However, work on code-based app quality evaluation for recommending high quality apps is rather limited. We compare our work with DroidVisor \cite{RustgiFRM17} which a popular recommendation system mainly based on user-generated information. DroidVisor is proposed to recommend apps based on similarity (app description), popularity (downloads), security (permissions) and usability (ratings). %Its ranking part is similar to our scenario.
	DroidVisor requires users to set four weights. We set the weight of similarity to 0 since our experiments are already conducted in the same app category. The weight of popularity, rating and security are all set to 1, which makes DroidVisor give higher score to popular and secure apps. The test apps are divided into three groups based on their scores as predicted labels. As shown in Fig.~\ref{compare}, the accuracy of AppQ is much higher than DroidVisor, especially in category News and Weather, indicating AppQ can overcome the disadvantage of missing feedback and better profile apps' real quality.
	
	\begin{figure}[htb]
		\centering
		\includegraphics[width=0.48\textwidth]{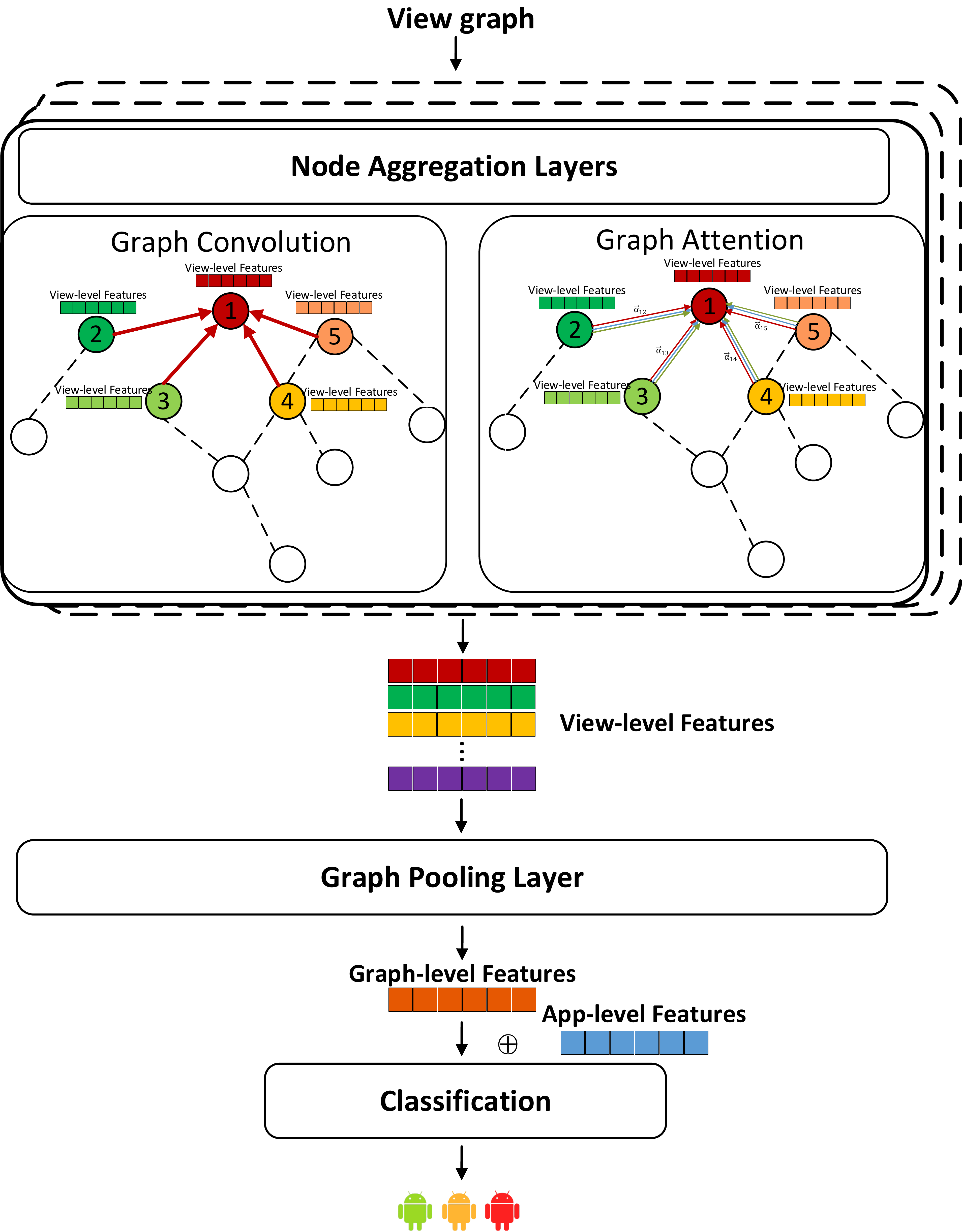}
		\caption{Graph-level representation learning with GCN or GAT}\label{GCN_GAT}
	\end{figure} 
	
	To demonstrate the effectiveness of our graph-to-vector method, we compare with two representative graph neural network models. We apply GCN \cite{KipfW17} and GAT \cite{GAT} to extract the app representation vectors from view graphs. These graph neural network models are initially proposed to solve node classification problems, with which graph pooling\cite{wu2020comprehensive,zhou2018graph} operations are integrated to generate higher graph-level representation based on node representations. As from Fig.~\ref{GCN_GAT}, after we get the attributed view graphs and app-level features, we feed them into node aggregation layers to get the view-level features for each view. In each node aggregation layer, every node will aggregate feature information from its 1-hop neighbours through the links. After several node aggregation operations, we will get the aggregated view-level feature vectors for each view that contain the feature information of the views on the graph and the structure information of the graph. Then the aggregated view-level features are input to the graph pooling layer. In the graph pool layer, the hidden feature vectors of each node are fed through a fully-connected layer, and then the mean pool operation is applied to the set of nodes to get the graph-level features. After that, the graph-level features are combined with app-level features to produce the final app-level features which incorporate both aggregated view-level and original app-level features. Got the new app-level features, we will follow the same classification process as above to classify the apps to different categories.
	
	On a part of the categories, AppQ performs better than the variants that replace the graph-to-vector module by GCN and GAT, while performing on par with them on other categories. It provides a strong proof that top-$k$ vertices and their neighbors are capable to represent the view graph. It is because even high quality apps always have several simple-designed views which users seldom use, e.g., account setting. AppQ only retain the features of main views that interact most closely with users, while filtering out unimportant views which are noises to the model.

	\subsection{Case study}
\begin{figure}[htb]
		\centering
		\includegraphics[width=.48\textwidth]{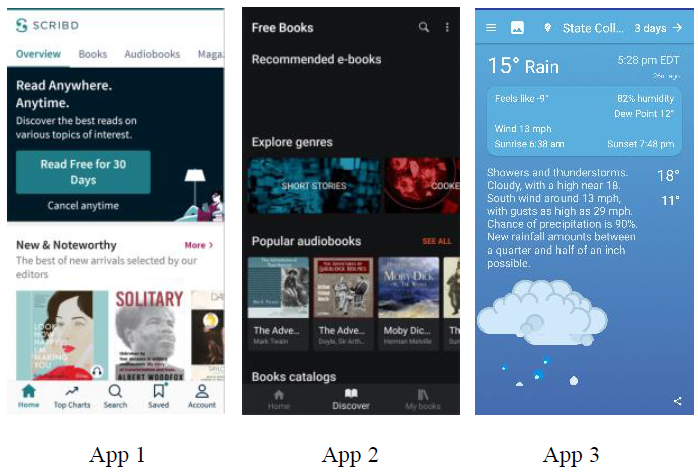}
		\caption{Case study}\label{f8}
	\end{figure}

	AppQ is able to %overcome the drawback of user generated statistics % to accurately evaluate the app based on its inborn features
	estimate an app's quality despite the limited users' feedback. As a case study, we test a few apps manually. As shown in Fig. \ref{f8} and Table \ref{t8}, app 1 is a popular online eBook with exquisite views, and it has been classified correctly. App 2 is also an eBook app whose design is much alike app 1 and the user experience is also good, which AppQ predicated level 3 as its quality. But due to the fact that it is a newly published app, it suffers from the cold-start problem and has not been well introduced to users. The same situation happens to app 3, which is a weather app. It is user friendly and there are no ads in its views; thus, AppQ regards it as a high quality app. However, it had only 48 ratings and 20 comments on January 7th, 2019. We searched its information again on February 26th, there was no significant change in downloads. The number of ratings and comments were slightly increased to 72 and 36, indicating it did not attract much attention from users. Therefore, there exists a discrepancy between the user-generated information and app's real quality. Purely based on user generated information to rank apps is not reliable. The cold-start problem hinders new released high quality apps from receiving due attention. With the analysis result from AppQ, app stores may give such apps a warm-start in the early stage. 
	
	\begin{table}[htbp]
		\caption{Apps in case study}
		\begin{center}
			\begin{tabular}{llll}
				\toprule  
				&App1&App2&App3\\
				\midrule  
				App name&Scribd&Free books&All clear weather\\
				Downloads&10,000,000&1,000&1,000\\
				Ratings&4.4&4.3&4.6\\
				Number of ratings&448,378&21&48\\
				Number of reviews&104,747&5&20\\
				Predicted label&3&3&3\\
				\bottomrule 
			\end{tabular}
			\label{t8}
		\end{center}
	\end{table}
	
	\subsection{Limitations and Discussions}
	Many factors can affect the rating of an app. For example, whether the function meets individual users' expectations or whether users like the theme or content of the app. However, these factors are sometimes very subjective, and that is why the same app may receive all kinds of ratings, from 1 star to 5 stars from different people. These factors cannot be measured by AppQ directly. Note that although AppQ has mainly focused on UI to estimate app quality, our experiments have demonstrated its effectiveness and efficiency. 
	
	One potential application of AppQ is to detect promoted low-quality apps, which may be downloaded in a great number in a short time and given very high rating scores. With AppQ, such apps will less likely to be ranked high even they have lots of downloads and getting mostly five stars.  
	
	\section{CONCLUSION}
	Motivated to overcome the drawbacks of current ranking strategies and to give newly published apps a warm-start, we propose AppQ to grade apps based on their inborn code-level features. AppQ performs code analysis to extract app-level features and dynamically runs apps to capture the switching among views. It constructs an attributed view graph that is encoded by a sequence of key vertices and their neighbors to highlight the importance of main views within an app. The view graph is further converted to a feature vector. An ensemble of machine learning approaches is then applied to evaluate and classify apps into different quality categories. The evaluation results demonstrates the effectiveness and efficiency of AppQ.
	
	In our future work, we are planning to explore more effective graph-to-vector approaches to better profile relations between apps' quality and UI. Exploring more features for characterizing apps' structures is also being investigated.

	\bibliographystyle{ieeetr}
	\vspace{3cm}
	\bibliography{ref}

\end{document}